\documentclass[12pt,a4paper]{article}
\usepackage{bbm}
\usepackage{amsmath}
\usepackage{amssymb}
\usepackage{hyperref}
\usepackage[mathcal]{euscript}

\newcounter{thMM}[section]
\newcounter{leMM}[section]
\newcounter{deFF}[section]
\newcounter{exMP}[section]
\newcounter{prOP}[section]
\newcounter{coRR}[section]
\newcounter{coexMP}[section]
\newcounter{remAR}[section]
\newenvironment{definition}[1][Definition]{\refstepcounter{deFF}\trivlist
   \item[{\bf #1~\thesection.\arabic{deFF}.}]\rm\hskip3pt}{\endtrivlist}

\newenvironment{proposition}[1][Proposition]{\refstepcounter{prOP}\trivlist
   \item[{\bf #1~\thesection.\arabic{prOP}.}]\it\hskip3pt}{\endtrivlist}
\newenvironment{corollary}[1][Corollary]{\refstepcounter{coRR}\trivlist
   \item[{\bf #1~\thesection.\arabic{coRR}.}]\it\hskip3pt}{\endtrivlist}

\newenvironment{remark}[1][Remark]{\refstepcounter{remAR}\trivlist
\item[{\bf #1~\thesection.\arabic{remAR}.}]\rm\hskip3pt}{\endtrivlist}

\newenvironment{proof}[1][Proof]{\begin{trivlist}
\item[\hskip \labelsep {\bfseries #1}]}{\end{trivlist}}

\newcommand{\proofend}{\flushright $\square$}

\newcommand{\D}{\mathcal{D}}

\DeclareMathOperator{\Vect}{Vect}

\DeclareMathOperator{\Span}{Span}

\makeatletter
\renewcommand\@biblabel[1]{#1.}
\makeatother

\begin{document}
\author{Andrew James Bruce  \\ \small \emph{UBIC, The University of Brighton},\\\small  \emph{Brighton, BN2 4GJ, UK}\\
 \small{\emph{email:} \texttt{andrewjamesbruce@googlemail.com}  } }
\date{\today}
\title{Contact structures and supersymmetric  mechanics}
\maketitle

\begin{abstract}
We review  the  relationship between contact structures on supermanifolds and supersymmetric mechanics in the superspace formulation. This allows one to use the language of contact geometry when dealing with the $d=1$, $N=2$ super-Poincar\'e algebra.
\end{abstract}
\section{Introduction}

In this work we reexamine the  contact form associated with $N=2$ supersymmetric mechanics, or rather the superspace realisation of the  $d=1$, $N=2$ super-Poincar\'e algebra.    This allows the interpretation of the SUSY transformations in superspace as  strict contactomorphisms. We  show how this contact form on $\mathbbmss{R}^{1|2}$ can be understood in terms of more established coset space methods and the appropriate Maurer--Cartan form. To the authors knowledge, the explicit link between coset space methods and contact geometry has not been discussed in the literature before. The initial link between supersymmetry and contact structures was established by Manin \cite{Manin1991} and was further explored by  Schwarz  and his collaborators \cite{Schwarz1992}.  Much of the work presented here should be considered as a review of established, if not well-known ideas. \\

Recently there has been renewed interest in contact structures on super and graded manifolds; for instance see Grabowski \cite{Grabowski2011} and Mehta \cite{Mehta2011}. One interesting non-classical feature of contact structures on supermanifolds is that one has both \emph{even} and \emph{odd} structures.  In this work we concentrate on a very specific even contact structure and how it arises in the context of supersymmetry. The constructions in this work are model independent, that is with no reference to some super-action. For reviews of how to construct actions for supersymmetric mechanics see  \cite{Bellucci2006,deLimaRodrigues2002,deLimaRodrigues2002B}. \\

With the relation between SUSY and contact structures being our primary goal here, let us present a lightning review of classical contact structures highlighting the elements we need later. Recall that a precontact structure on a manifold is a one-form that is nowhere vanishing\footnote{We assume all structures to be global and skip questions of orientability.}. Associated with every precontact structure on a manifold is a hyperplane distribution, that is  a subbundle of the tangent bundle of corank $1$. The hyperplane distribution is defined to be the kernel of the precontact structure.  That is if we denote the precontact structure as $\alpha \in \Omega^{1}(M)$ then $\D_{\alpha} = \ker{\alpha}$. That is the hyperplane distribution consists of all vector fields $X \in \Vect(M)$ such that $i_{X}\alpha =0$.\\

A contact structure on a manifold of dimension $(2n +1)$ ($n \in \mathbbmss{N}^{*})$ is a precontact structure with the extra requirement that the  exterior derivative of the structure  is  non-degenerate on the associated hyperplane distribution.  That is there are no non-zero vector fields $X \in \D_{\alpha}$ such that $i_{X}(d \alpha)=0$.\\

Let $(M, \alpha)$ be a contact manifold. A diffeomorphism $\phi: M \rightarrow M$  is said to be a \emph{contactomorphism} if and only if $\phi^{*}\alpha = f \alpha$ for some nowhere vanishing function $f \in C^{\infty}(M)$. A diffeomorphism $\phi: M \rightarrow M$  is said to be a \emph{strict contactomorphism} if and only if $\phi^{*}\alpha = \alpha$.  Contactomorphisms preserve hyperplane distributions.  A vector field $X \in \Vect(M)$ is said to be a \emph{contact vector field} if and only of $L_{X}\alpha = f \alpha$. If $f = 0$, then then vector field is said to be a \emph{strict contact vector field}.\\

The literature on  contact structures is vast and is constantly growing. A good description of contact structures on supermanifolds and their idiocrasies can be found in \cite{Grabowski2011}.  We do not employ anything deep from the general theory of contact structures and so direct the reader that is unfamiliar with classical contact structures to introductory texts. For example see Appendix 4 of Arnold's book \cite{Arnold1989}. For preliminaries on supermanifolds  we recommend \cite{Voronov1992}.

\section{$N=2$ SUSY mechanics in superspace}

Consider the superspace $\mathbbmss{R}^{1|2}$ equipped with local coordinates $(t, \theta, \overline{\theta})$. The SUSY transformations are \emph{defined} in this superspace to be

\begin{eqnarray}\label{n2susytransform}
t &\rightarrow& t' = t+ i \left(\epsilon \bar{\theta} - \theta \bar{\epsilon}  \right),\\
\nonumber \theta &\rightarrow& \theta' = \theta + \epsilon,\\
\nonumber \bar{\theta} &\rightarrow& \bar{\theta}' = \bar{\theta}+ \bar{\epsilon},
\end{eqnarray}

where $\epsilon$ and $\bar{\epsilon}$ real Grassmann odd parameters. The factor of $i = \sqrt{-1}$ is included to ensure that the product of two real Grassmann odd objects is real. Thus, the real nature of time is maintained.\\

An \textbf{even superfield} is  an even real function on $\mathbbmss{R}^{1|2}$. Expanded out in components we have

\begin{equation}\nonumber
\Phi(t, \theta, \overline{\theta}) = q(t) + i \theta \psi(t) + i \bar{\psi}(t) \bar{\theta} + i \theta \bar{\theta} b(t),
\end{equation}

where $q(t)$ and $b(t)$ are even pure real functions of time. The components  $\psi(t)$ and  $\bar{\psi}(t)$ are odd in nature.  To include odd functions one requires the use of external odd parameters or constants which can be employed as Fourier coefficients in defining odd functions of even variables. We will not dwell on this.  A general even superfield describes the $(\mathbf{1},\mathbf{2},\mathbf{1})$ supermultiplet. Other supermultiplets exist: the $(\mathbf{2},\mathbf{2},\mathbf{0})$ supermultiplet is described by a complex even chiral superfield and the $(\mathbf{0},\mathbf{2},\mathbf{2})$ by an odd chiral superfield. For details see \cite{Bellucci2006}.   We will generally not be working with supermultiplets and will have no course to consider odd superfields in any detail. \\

Let us  as standard introduce the two vector fields

\begin{equation}
Q =  \frac{\partial}{\partial \theta} + i \bar{\theta} \frac{\partial}{\partial t} \hspace{15pt}\textnormal{and}\hspace{15pt}\bar{Q} = \frac{\partial}{\partial \bar{\theta}} + i \theta \frac{\partial}{\partial t},
\end{equation}

as the vector fields that ``implement" the SUSY transformations viz

\begin{equation}\nonumber
\delta \Phi =\left( \epsilon Q + \bar{\epsilon} \bar{Q}\right)[\Phi] =  \delta t \frac{\partial \Phi}{\partial t} + \delta \theta \frac{\partial \Phi}{\partial \theta} + \delta \bar{\theta} \frac{\partial \Phi}{ \partial \bar{\theta}}.
\end{equation}

It is straight forward to see that in terms of the $(\mathbf{1},\mathbf{2},\mathbf{1})$ supermultiplet\\
\begin{center}
\begin{tabular}{ll}
$\delta q(t) = i \epsilon \psi + i \bar{\psi} \bar{\epsilon}$, &  $\delta \psi(t)  = (b - \dot{q}) \bar{\epsilon}$,\\
$\delta \bar{\psi}(t) = \epsilon (b+ \dot{q})$,  & $\delta b(t) = i \dot{\bar{\psi}} \bar{\epsilon} - i \epsilon \dot{\psi}$.
\end{tabular}
\end{center}

It is also easy to show that the graded commutator

\begin{equation}
[Q,\bar{Q}] =   Q \circ \bar{Q} + \bar{Q} \circ Q = 2 i \frac{\partial}{\partial t},
\end{equation}

and that all other (graded) commutators involving $Q, \bar{Q}$ and $\frac{\partial}{\partial t}$ are identically zero. \\

Up to this point our presentation of the $d=1$, $N=2$  super-Poincar\'e algebra has been rather standard. We now wish to introduce a geometric structure very similar to a classical contact structure and then interpret as much as possible in terms of (even) contact supergeometry. \\

\begin{definition}
The \textbf{super contact form} on $\mathbbmss{R}^{1|2}$ is defined to be the Grassmann odd  one-form
 \begin{equation}
  \alpha = dt + i\left(\theta d\bar{\theta} +  \bar{\theta} d \theta\right).
  \end{equation}
\end{definition}

\begin{remark}
The space $\mathbbmss{R}^{3}$, with local coordinates $(x,y,z)$ can be equipped with the contact form $\alpha :=   dz \pm x dy \mp yd x $. Thus we can consider the super contact form on $\mathbbmss{R}^{1|2}$ to be a natural ``superisation" of this contact form on $\mathbbmss{R}^{3}$. Note that the contact form on  $\mathbbmss{R}^{3}$ as given above is not the \emph{standard contact structure} which is given by $dz + xdy$.
\end{remark}
We define differential (pseudo)forms on supermanifold to be superfunctions on the total space of the antitangent bundle. For the case at hand we have $\Pi T(\mathbbmss{R}^{1|2})$ which we equip with fibre coordinates $(dt, d \theta, d\bar{\theta})$. Here the coordinate $dt$ is  \emph{odd} as where  the coordinates $d\theta$ and $d\bar{\theta}$ are \emph{even}. General coordinate changes on $\mathbbmss{R}^{1|2}$ are of the form $t \rightarrow t' = t'(t, \theta, \bar{\theta})$, $\theta \rightarrow \theta' = \theta'(t, \theta, \bar{\theta})$ and $\bar{\theta} \rightarrow \bar{\theta}' = \bar{\theta}'(t, \theta, \bar{\theta})$. These changes of coordinates induce vector bundle automorphisms of the form

\begin{eqnarray}\label{bundleautomorphisms}
dt' &=& dt\frac{\partial t'}{\partial t} + d\theta \frac{\partial t'}{\partial \theta} + d \bar{\theta}\frac{\partial t'}{\partial \bar{\theta}},\\
\nonumber d\theta' &=&  dt \frac{\partial \theta'}{\partial t} + d\theta \frac{\partial \theta'}{\partial \theta} + d\bar{\theta} \frac{\partial \theta'}{\partial \bar{\theta}}, \\
 \nonumber d\bar{\theta}' &=&  dt \frac{\partial \bar{\theta}'}{\partial t} + d\theta \frac{\partial \bar{\theta}'}{\partial \theta} + d\bar{\theta} \frac{\partial \bar{\theta}'}{\partial \bar{\theta}}.
\end{eqnarray}

 A one-form on a supermanifold is then a function on the respective anticotangent bundle linear in fibre coordinates. The exterior derivative acting on differential (pseudo)forms on $\mathbbmss{R}^{1|2}$ is the homological vector field
  \begin{equation}\nonumber
  d = dt \frac{\partial}{\partial t} + d \theta \frac{\partial}{\partial \theta} + d \bar{\theta} \frac{\partial}{\partial \bar{\theta}}.
 \end{equation}

 By homological one means that $[d,d] = 2 d^{2}=0$ where the bracket is the graded Lie bracket.\\

The interior derivative also naturally generalises to supermanifolds. A vector fields on $\mathbbmss{R}^{1|2}$ is of the form

 \begin{equation}\nonumber
 X = X_{t}(t,\theta, \overline{\theta})\frac{\partial}{\partial t} + X_{\theta}(t, \theta, \bar{\theta})\frac{\partial}{\partial \theta} + X_{\bar{\theta}}(t, \theta, \bar{\theta})\frac{\partial}{\partial \bar{\theta}},
 \end{equation}

  in obvious notation. The interior derivative with respect to a vector field $X \in \Vect(\mathbbmss{R}^{1|2})$ (take to be homogeneous in parity) is given by a vector field $i_{X} \in \Vect(\Pi T(\mathbbmss{R}^{1|2}))$  which in local coordinates is

  \begin{equation}\nonumber
  i_{X} = (-1)^{\widetilde{X}} \left(X_{t}\frac{\partial}{\partial dt} + X_{\theta}\frac{\partial}{\partial d\theta} + X_{\bar{\theta}}\frac{\partial}{\partial d\bar{\theta}}\right  ),
   \end{equation}

   where  $\widetilde{X}$ is the Grassmann parity of the vector field. Extension to inhomogeneous vectors fields is via linearity. The Lie derivative also generalises via $L_{X} = [d, i_{X}]$. \\

The claim is that the super contact form is a genuine contact structure on the supermanifold $\mathbbmss{R}^{1|2}$. In particular  $\alpha$ is a Grassmann odd one-form that is non-vanishing in the sense that $\alpha|_{\theta = 0, \bar{\theta}=0} \neq 0$. As such it defines a corank $(1|0)$ hyperplane distribution via its kernel.

\begin{itemize}
\item The hyperplane distribution is defined as
\begin{equation}\nonumber
\D_{\alpha} := \ker(\alpha) = \Span\left\{ \frac{\partial}{\partial \theta} - i \bar{\theta} \frac{\partial }{\partial t},\frac{\partial}{\partial \bar{\theta}} - i \theta \frac{\partial }{\partial t}  \right\} \subset T(\mathbbmss{R}^{1|2}).
\end{equation}
We see that it consists of the span of two odd vector fields. Thus the corank is indeed $(1|0)$, ie. it consists of one less even vector field in its basis as compared to the tangent bundle.  Note that these odd vector fields are the standard SUSY covariant derivatives, which are commonly introduced in response to $\frac{\partial \Phi}{\partial \theta}$  and $\frac{\partial \Phi}{\partial \bar{\theta}}$ not transforming as  superfields.  Let us denote these vector fields as
\begin{equation}
\mathbbmss{D} = \frac{\partial}{\partial \theta} - i \bar{\theta} \frac{\partial }{\partial t} \hspace{15pt}\textnormal{and} \hspace{15pt} \bar{\mathbbmss{D}} = \frac{\partial}{\partial \bar{\theta}} - i \theta \frac{\partial }{\partial t}.
\end{equation}
\item One also has to check the non-degeneracy condition on $\D_{\alpha}$. First note $ \omega = d \alpha =   2 i d\theta d \bar{\theta}$ is  an even symplectic structure on $\mathbbmss{R}^{0|2}$. Via direct calculation
\begin{equation}\nonumber
i_{\mathbbmss{D}}(d \alpha) = -2i d\bar{\theta}  \hspace{15pt}\textnormal{and}\hspace{15pt} i_{\bar{\mathbbmss{D}}}(d \alpha) = -2i d\theta,
\end{equation}
which implies the non-degeneracy condition. That is there are no non-zero vector fields in $\D_{\alpha}$ that annihilate the two-form $ \omega = d\alpha$.
\end{itemize}

Following Manin \cite{Manin1991}, we  refer to the hyperplane distribution $\D_{\alpha}$ as the $\mathbf{\textnormal{\textbf{SUSY}}_{2}}$ \textbf{structure}.\\

One can always (at least locally) associate a precontact structure with \emph{any}  hyperplane distribution. The remarkable point is that the structure associated with supersymmetry  is in fact contact.   \\

\begin{proposition}
The super contact form $\alpha = dt + i\left(\theta d\bar{\theta} +  \bar{\theta} d \theta\right)$ on $\mathbbmss{R}^{1|2}$ is invariant under SUSY transformations (\ref{n2susytransform}).
\end{proposition}

\begin{proof}
Via direct computation we see that under the SUSY transformations
\begin{equation}
\nonumber dt' = dt- i \bar{\epsilon} d\theta - i \epsilon d \bar{\theta},\hspace{20pt} d\theta' = d \theta, \hspace{20pt} d \bar{\theta}' =  d \bar{\theta},
\end{equation}
\newpage
where hence
\begin{equation}
\nonumber \alpha' = dt- i \bar{\epsilon} d\theta - i \epsilon d \bar{\theta} + i(\theta + \epsilon) d \bar{\theta} + i(\bar{\theta} + \bar{\epsilon})d \theta = \alpha.
\end{equation}
\proofend
\end{proof}

This implies that the SUSY transformations preserve the hyperplane distribution $\D_{\alpha}$.

\begin{corollary}
 The vector fields $Q$ and $\bar{Q}$ are strict contact vector fields of the the super contact form, i.e.
 \begin{equation}\nonumber
 L_{Q}\alpha = 0,   \hspace{15pt}\textnormal{and} \hspace{15pt} L_{\bar{Q}}\alpha=0.
 \end{equation}
\end{corollary}

Of course the vector fields $Q$ and $\bar{Q}$  represent infinitesimal strict contactomorphisms. This in turn implies that
\begin{center}
\begin{tabular}{ll}
$[Q,\mathbbmss{D}] = 0$, &  $[\bar{Q}, \bar{\mathbbmss{D}}] =0$,\\
$[Q, \bar{\mathbbmss{D}}] =0$ & $[\bar{Q}, \mathbbmss{D}] =0$,
\end{tabular}
\end{center}

as expected. Direct computation shows that

\begin{equation}
\nonumber [\mathbbmss{D},\bar{\mathbbmss{D}}] = - 2 i \frac{\partial}{\partial t} \not \in \D_{\alpha},
\end{equation}

and thus as expected $\D_{\alpha}$ is not involutive in the sense of Frobenius.\\

From the classical theory of contact structures, we know that there is a privileged strict contact  vector field known as the \textbf{Reeb vector field}. We will denote the Reeb vector field by $P$, the reason why will become clear.  This vector field is defined uniquely by the conditions

\begin{equation}\nonumber
i_{P}\alpha = 1 \hspace{15pt}\textnormal{and} \hspace{15pt} i_{P}(d\alpha) =0.
\end{equation}

\begin{proposition}
On $\mathbbmss{R}^{1|2}$ equipped with the super contact  structure $\alpha = dt + i\left(\theta d\bar{\theta} +  \bar{\theta} d \theta\right) $ the Reeb vector field is given by $P = \frac{\partial }{\partial t}$.
\end{proposition}

\begin{proof}
Via direct computation:
\begin{equation}\nonumber
i_{\frac{\partial}{\partial t }}\left( dt + i \bar{\theta} d\theta  + i \theta d \bar{\theta}\right) = 1 \hspace{15pt}\textnormal{and} \hspace{15pt} i_{\frac{\partial}{\partial t}}\left(-2i d\theta d \bar{\theta} \right)=0.
\end{equation}
\proofend
\end{proof}

Thus we see that the  Reeb vector field corresponds to temporal translations. We are then led to an interesting interpretation of the  so-called $N=2$ \emph{right supertranslation and time-translation algebra}:

\begin{eqnarray}
[Q,\bar{Q}] &=& 2 i P,\\
\nonumber [Q, P] &=& [\bar{Q}, P] = 0,
\end{eqnarray}

as a  Lie subalgebra of the Lie algebra of  strict contact vector fields of the super contact structure.\\

\begin{remark}
The super contact structure is also invariant under $R$-transformations;
\begin{center}
\begin{tabular}{ll}
$\theta \rightarrow \theta' = e^{-i \beta} \theta $, & $\bar{\theta} \rightarrow \bar{\theta}' = e^{i \beta} \bar{\theta} $.
\end{tabular}
\end{center}
Infinitesimally $R$-transformations can be ``implemented" by the even vector field
\begin{equation}\nonumber
R = - i \left(\theta \frac{\partial}{\partial \theta} - \bar{\theta} \frac{\partial}{\partial \bar{\theta}} \right).
\end{equation}
Direct computation gives

\begin{center}
\begin{tabular}{ll}
$[R,Q] = i Q$,   &   $[R,\bar{Q}] = -i \bar{Q}$,\\
$[R,P] =0$ & $[R,R]=0$.
\end{tabular}
\end{center}
In short, $R$-symmetry can also be understood in terms of the Lie algebra of strict contact vector fields.
\end{remark}

Hamiltonian vector fields play an important role in both symplectic and contact geometry. In particular they are important  as they represent infinitesimal symmetries.

\begin{definition}
Let $\Upsilon(t, \theta, \bar{\theta}) = a(t) + i \theta \chi(t) + i \bar{\chi}(t)\bar{\theta} + i \theta \bar{\theta}c(t)$ be an even, but otherwise arbitrary   superfield. The associated \textbf{Hamiltonian vector field} is the unique (Grassmann even) vector field $X_{\Upsilon} \in \Vect(\mathbbmss{R}^{1|2})$ that satisfies

\begin{equation}\nonumber
i_{X_{\Upsilon}}\alpha = \Upsilon,  \hspace{20pt}\textnormal{and}\hspace{20pt} i_{X_{\Upsilon}}(d\alpha) = P(\Upsilon)\alpha - d \Upsilon.
\end{equation}
\end{definition}

\begin{remark}
Note that we only consider the Hamiltonian vector field associated with a Grassmann \emph{even} superfield. The analogous definition for \emph{odd} superfields will contain extra sign factors. We will have no need to consider Grassmann odd superfields as they \emph{cannot} generate contact Hamiltonian vector fields. In particular the Grassmann parity of the SUSY contact structure would not be preserved.
\end{remark}

In local coordinates the Hamiltonian vector associated with $\Upsilon$ is given by

\begin{eqnarray}
X_{\Upsilon} &=& \left( a(t) + \frac{i}{2} \left( \theta \chi(t) + \bar{\chi}(t)\bar{\theta} \right) \right)\frac{\partial}{\partial t}\\
\nonumber &+& \frac{i}{2}\left( \bar{\mathbbmss{D}}\Upsilon \right)\frac{\partial}{\partial \theta} + \frac{i}{2}\left( \mathbbmss{D}\Upsilon \right)\frac{\partial}{\partial \bar{\theta}}.
\end{eqnarray}

Note that Hamiltonian vector fields are contact vector fields, but are not in general strict contact vector fields. For the case at hand we see that
\begin{equation}\nonumber
L_{X_{\Upsilon}}\alpha = \dot{\Upsilon} \alpha,
\end{equation}
where we have used ``dot" to denote the time derivative. Clearly superfields that are constant in time generate strict contactomorphisms. In order to define a contactomorphism that is not strict the  (even) superfield $\dot{\Upsilon}$ must be nowhere vanishing in the sense that $\dot{\Upsilon}|_{\theta = \bar{\theta}=0} = \dot{a}(t) \neq 0$  anywhere on $\mathbbmss{R}$.   \\

Written out explicitly the infinitesimal contactomorphisms associated with the Hamiltonian vector field are of the form

\begin{eqnarray}
\delta t &=& a(t) + \frac{i}{2}\left( \theta \chi(t) + \bar{\chi}(t) \bar{\theta} \right),\\
\nonumber \delta \theta &=& \frac{1}{2} \left( \bar{\chi}(t) + \theta(\dot{a}(t) +c(t)) - i \theta \bar{\theta}\dot{\bar{\chi}}(t) \right),\\
\nonumber \delta \bar{\theta} &=& - \frac{1}{2}\left( \chi(t) - (\dot{a}(t)- c(t)) \bar{\theta} + i \dot{\chi}(t) \theta \bar{\theta} \right).
\end{eqnarray}

As already established one can consider the (infinitesimal) SUSY transformations as strict contactomorphisms of the super contact form. Moreover, we can now interpret the SUSY transformations as being generated by the Hamiltonian vector field associated with the time independent superfield

\begin{equation}
\Upsilon(\theta, \bar{\theta}) := 2 \left(\epsilon \bar{\theta}  - \theta \bar{\epsilon}\right).
\end{equation}

Another interesting transformation is generated by the superfield

\begin{equation}\nonumber
\Upsilon(t, \theta, \bar{\theta}) := \lambda t + 2 t \left( \epsilon \bar{\theta} - \theta \bar{\epsilon} \right),
\end{equation}

here $\lambda$ is a real parameter.  Note that $\dot{\Upsilon}|_{\theta = \bar{\theta}=0} \neq 0$ assuming that $\lambda$ is nonzero. The associated transformations  are given by

\begin{eqnarray}
\delta t &=&  \lambda t + i t \left( \epsilon \bar{\theta} - \theta \bar{\epsilon} \right),\\
\nonumber \delta \theta &=& \frac{\lambda}{2} \theta + \epsilon \left(t - i \theta \bar{\theta}  \right),\\
\nonumber  \delta \bar{\theta} &=& \frac{\lambda}{2} \bar{\theta} + \left( t+ i \theta \bar{\theta} \right)\bar{\epsilon}.
\end{eqnarray}

Notice that the above transformations are ``superconformal-like". See for example \cite{Ivanov2004} for details about conformal and superconformal mechanics.\\

If desired, one can now calculate transformations generated by Hamiltonian vector fields at the level of the various supermultiplets. Details are left to the reader.

\section{The relation between the super contact form and coset space methods.}

The standard approach to constructing SUSY covariant derivatives is to employ the methods of homogeneous spaces applied to supermanifolds. One constructs the appropriate Maurer--Cartan form  and then extracts the covariant derivatives. For supersymmetric mechanics this method is rather involved considering one could straightforwardly \emph{guess} the SUSY covariant derivatives. However, the coset method is rather general and applies to far more complicated theories. An accessible review to the construction of the Maurer--Cartan form associated with $\mathcal{N}=1$ supersymmetric field theory can be found in \cite{Bagger1996}. For a review of the methods as applied to supersymmetric mechanics and superconformal mechanics see \cite{Bellucci2006}. We will draw heavily from both these works in this section and direct the reader to them for details. \\

We will now outline the construction of the Maurer--Cartan form associated with $N=2$ supersymmetric mechanics and relate this to the super contact structure. Unsurprisingly the two structures are closely linked.  \\

Let $G$ be the supergroup generated by the SUSY transformations and let $H$ be the group generated by  temporal translations. The group $H$ is a stabiliser subgroup of $G$.  As the supergroup $G$ acts on $\mathbbmss{R}^{1|2}$ transitively we have:\\

\parbox[h][50pt][l]{420pt}{
\emph{The action of the supergroup $G$ can be realised by the left multiplication on the coset $G/H$ and the coordinates which parameterise the coset are given by the coordinates on $\mathbbmss{R}^{1|2}$}.}\\

Then

\begin{equation}
g = e^{i \left( t \frac{\partial}{\partial t} + \theta Q + \bar{\theta}\bar{Q}  \right)},
\end{equation}

is a natural parametrisation of the coset $G/H$.\\

\begin{definition}
The  (left invariant) \textbf{Maurer--Cartan form} is the  one form  given by
\begin{equation}\nonumber
i \Omega = g^{-1}\left(dg  \right).
\end{equation}
\end{definition}

We include an overall factor of ``$i$" for convenience.  It is natural to consider the Maurer--Cartan form as a tangent bundle valued one-form on  $\mathbbmss{R}^{1|2}$. To calculate this one appeals to \emph{the Hadamard lemma}, which of course is closely related to the \emph{Baker--Campbell--Hausdorff formula}:

\begin{eqnarray}
\nonumber i \Omega &=& e^{-i \left( t \frac{\partial}{\partial t} + \theta Q + \bar{\theta}\bar{Q}  \right)}\left(  de^{i \left( t \frac{\partial}{\partial t} + \theta Q + \bar{\theta}\bar{Q}  \right)} \right)\\
\nonumber &=& i \left( (dt + i(\theta d\bar{\theta} + d \theta \bar{\theta}) ) \frac{\partial }{\partial t}  + d \theta Q + d \bar{\theta} \bar{Q}\right).
\end{eqnarray}

 \noindent \textbf{Statement:} \emph{The Maurer--Cartan form ``contains" the super contact structure as the component belonging to the stability subgroup generated by temporal translations.}\\

  It is well know that this component transforms as a  connection (see \cite{Bellucci2006}) and thus can be used to construct the covariant derivatives. Thus  the Maurer--Cartan form and super contact structure on $\mathbbmss{R}^{1|2}$ are very closely related.

\section{$N=1$ SUSY mechanics in superspace}
An analogous interpretation  of  the covariant derivative and the SUSY algebra associated with $N=1$ supersymmetric mechanics also exists. For brevity we just outline the constructions.  The superspace relevant here is $\mathbbmss{R}^{1|1}$, which we equip with local coordinates $(t,\theta)$. The super  contact structure in this case is given by

\begin{equation}
\alpha = dt + i \theta d\theta.
\end{equation}

This super contact form is invariant under the SUSY transformations

\begin{equation}
t \rightarrow t' = t + i \epsilon \theta \hspace{15pt}\textnormal{and} \hspace{15pt} \theta \rightarrow \theta' = \theta + \epsilon.
\end{equation}

Associated with these transformations is the vector field defined by $\delta \Phi = \epsilon Q[\Phi]$ for any superfield on $\mathbbmss{R}^{1|1}$:
\begin{equation}
Q = \frac{\partial }{\partial \theta} + i \theta \frac{\partial }{\partial t}.
\end{equation}

The hyperplane distribution associated with $\alpha$ is

\begin{equation}\nonumber
\D_{\alpha} = \Span\left \{ \frac{\partial }{\partial \theta} - i \theta \frac{\partial }{\partial t}   \right\},
\end{equation}

where we recognise the single odd basis vector to be the SUSY covariant derivative $\mathbbmss{D}$. Clearly we have a distribution of corank $(1|0)$.  The hyperplane distribution $\D_{\alpha}$ will be referred to as the $\mathbf{\textnormal{\textbf{SUSY}}_{1}}$ \textbf{structure} following Manin. The Reeb vector field  corresponds to temporal translations

\begin{equation}\nonumber
P = \frac{\partial }{\partial t}.
\end{equation}

The so-called $N=1$ \emph{right supertranslation and time-translation algebra}:

\begin{equation}
[Q,Q] = 2 i P,  \hspace{20pt} [Q,P] = 0,
\end{equation}

can be interpreted as a Lie subalgebra of the Lie algebra of strict contact vector fields.

\section{Concluding remarks}
We have reexamined  how the $N=2$ SUSY algebra of supersymmetric mechanics can be understood in terms of an even contact  structure on the supermanifold $\mathbbmss{R}^{1|2}$. In particular:
\begin{enumerate}
\item The SUSY covariant derivatives $\mathbbmss{D}$  and $\bar{\mathbbmss{D}} \in \Vect(\mathbbmss{R}^{1|2})$ are understood to be a basis for the hyperplane distribution associated with the super contact structure.
    \item The $N=2$ SUSY algebra is understood in terms of the Lie algebra of ``strict contact vector fields" of the super contact structure.
    \item The  super contact structure is the piece of the Maurer--Cartan form associated with the stability subgroup of the supergroup generated by the SUSY transformations.
\end{enumerate}

The situation for $N=1$ supersymmetric mechanics was briefly outlined.\\

The case of extended supersymmetries can be done directly. From a geometric point of view extended supersymmetries may be more interesting. However, from a physics point of view superspace methods become clumsy for mechanics with a large number of supersymmetries. In particular superspace methods lead to reducible representations at the level of the supermultiplets. One has to introduce constraints and  construct irreducible representations that typically hold on-shell only.  For a review of the irreducible representations of extended supersymmetric mechanics see \cite{Toppan2006}.\\

The work presented here  is very  far from a complete study of  contact structures on supermanifolds. In particular supersymmetric mechanics appears to be closely related to \emph{even contact supergeometry}. An even contact structure is a \emph{Grassmann odd} one-form and thus describes a corank $(1|0)$ distribution. That is the distribution has one less \emph{even vector} in its span as compared to the tangent bundle.\\

Even contact structures also feature in analysing the projective geometry of supercircles  $S^{1|m}$ ($m=1,2$) and the super Schwarzian derivative. See for example Duval \& Michel \cite{Duval2008}, though the study of contact structures on supercircles was initiated much earlier in 1986 by Radul \cite{Radul1986}. Schwarz \cite{Schwarz1992} considers superconformal geometry to be a special case of complex contact geometry on supermanifolds and links this with superconformal and topological conformal field theories.  Clearly contact structures on low dimensional supermanifolds is of some continuing  mathematical interest.  \\

\noindent \textbf{Acknowledgements} \\
The author would like to thank Prof. J. M. Figueroa-O'Farrill and Dr. D. J. Miller  for their comments on an earlier version of this work. The author must also thank Dr. J. Michel  and Prof. F. Toppan for further comments that have helped improve this work.

\vfill
\begin{center}
Andrew James Bruce\\
\small{\emph{email:} \texttt{andrewjamesbruce@googlemail.com}  }
\end{center}
\end{document}